\documentclass[12pt, draftclsnofoot, onecolumn]{IEEEtran}
\usepackage{amsmath}
\usepackage{amsthm}
\usepackage{amsfonts}
\usepackage{amssymb}
\usepackage{mathrsfs}
\usepackage{mathdots}
\usepackage{multirow}
\usepackage{cases}
\usepackage{cite}
\usepackage{graphicx}
\usepackage{epstopdf}
\usepackage{mathrsfs}
\usepackage{subfigure}
\usepackage{epstopdf}
\usepackage{color}
\usepackage{bm}
\usepackage{caption}
\usepackage{booktabs}
\usepackage{stfloats}
\allowdisplaybreaks[1]
\theoremstyle{remark}

\begin{document}
\bibliographystyle{IEEEtran}

\title{Secure Short-Packet Communications at the Physical Layer for 5G and Beyond
\thanks{Chen Feng and Hui-Ming Wang (corresponding author) are with Xi'an Jiaotong University. (Email: {fengxjtu@163.com, xjbswhm@gmail.com}).
}
\author{Chen Feng, and~Hui-Ming Wang,~\IEEEmembership{Senior Member,~IEEE}
}
}
\maketitle
\vspace{-12mm}
\begin{abstract}
Short-packet communication is a key technology to support two emerging application scenarios in 5G and beyond 5G, massive machine type communication (mMTC) and ultra-reliable low latency communication (uRLLC), which are introduced to satisfy the broader communication requirements of potential applications such as the internet of vehicles and industrial internet of things (IoT). The sharp increase in privacy data in various IoT applications has made security issues more prominent. The typical upper-layer encryption mechanism could not fully address the security challenge considering the resource restriction of IoT terminals. In this article, we investigate secure short-packet communication from the perspective of physical layer security (PLS), which can be regarded as a promising security solution in 6G. Specifically, the state-of-the-art development of fundamental information theory of secure short-packet communications and corresponding performance evaluation criterion in fading channels are summarized. Then we review recent works, which investigate short-packet communication systems (CSs) in different communication scenarios or with different security strategies from the perspective of PLS. Finally, we give future research directions and challenges.
\end{abstract}

\section{Introduction}
\label{I}
In the past two decades, driven by high-resolution image, audio and video content transmission services for high-speed wireless data access anytime and anywhere,  traditional wireless CSs have pursued higher transmission rates and larger capacities. Recently, intelligent communication and sensing technologies are increasingly entering other fields such as industry, medicine, and transportation, promoting the rapid development of the IoT technology. Driven by potential IoT applications, apart from providing higher bandwidth services, 5G also introduces two new application scenarios, mMTC and URLLC, to deal with diverse demands of data transmission.

Currently, 5G communication standardization is entering the final stage, which has spurred interest in 6G in the research community. 6G is expected to support more diversified application scenarios, and the above-mentioned two scenarios will be further explored. One of the main challenges in providing services for mMTC and URLLC applications is that future CSs must support short-packet communications, which is significantly different from current CSs with long blocklength for high bandwidth \cite{durisi2016toward}. For many potential IoT applications in 5G and beyond, providing reliable data transmission is a mandatory requirement. However, communication with short packets will result in a severe reduction in channel coding gain, making it difficult to ensure communication reliability \cite{poor2019fundamentals}. 
\begin{figure}[!t]
\begin{center}
\includegraphics[width=3.5 in]{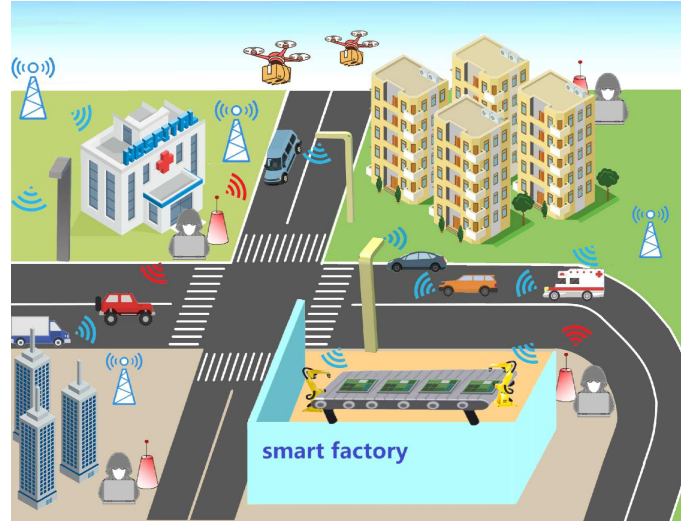}
\end{center}
\vspace{-5mm}
\caption{\small{Future IoT applications are facing unprecedented security challenges.}}
\label{fig1}
\vspace{-10mm}
\end{figure}

For many Mission-Critical IoT applications in 5G and beyond, they will be confronted with more severe secrecy challenges than other CSs because they often contain private data. As shown in Fig. \ref{fig1}, future IoT networks will be applied to internet of vehicles, home automation, intelligent logistics, smart factory, telemedicine and other aspects. While facilitating production and life, it also brings an unprecedented risk of privacy leakage. Taking intelligent transportation as an example, if a message is monitored, private information such as user identity or location information may be exposed, which may cause serious consequences. In previous CSs, the upper-layer encryption mechanism based on cryptography technology is used to provide security guarantee for information transmission, which encounters objective challenges in IoT systems:
\begin{itemize}
\item
Encryption relies on the effective generation and distribution of keys, and requires a unified key management infrastructure such as base station and security server, which poses challenges to many IoT networks that lack fixed infrastructure. What's more, massive numbers of IoT terminals will form a highly dynamic heterogeneous network, making it difficult to perform unified key management, negotiation, distribution and update.
\end{itemize}

\begin{itemize}
\item
A communication mode characterized by numerous sporadic short packets is used in short-packet communications, in which the message only contains hundreds of bits. Traditional encryption scheme requires multiple signaling interactions to generate the pairing key, which will cause a lot of extra overhead, greatly reduce spectrum efficiency, and consume too much terminal resources and energy when deployed in IoT system.
\end{itemize}

\begin{itemize}
\item
The future IoT systems involve a large number of terminals that are usually simple in structure and resource-constrained for cost considerations, which cannot support complex signal processing.  IoT systems usually use lightweight encryption protocols to reduce the demand for resources at the expense of system reliability and security.
\end{itemize}

Distinct from the principle of upper-layer encryption mechanism, PLS aims at reinforcing the security of CSs by exploiting the imperfections in the physical layer of the protocol stack, such as noise, interference, and the variation of channel strength in wireless channels \cite{poor2019fundamentals}. The PLS technology can be regarded as promising secure solutions in 6G with following advantages:
\begin{itemize}
\item
PLS does not rely on key transmission, its security is independent of the computing power of the eavesdropping terminal, and its dependence on infrastructure is reduced.
\end{itemize}
\begin{itemize}
\item
PLS can provide a low-complexity security solution, which is suitable for resource-constrained IoT nodes and helps reduce communication delays.
\end{itemize}
\begin{itemize}
\item
PLS offers ``built-in" security that is generally agnostic to the network infrastructure, and provides better scalability as the size of an IoT network increases\cite{poor2019fundamentals}.
\end{itemize}

In the past decade, extensive research has been conducted on the system design and secrecy performance analysis of various infinite blocklength (IFB) CSs from the perspective of PLS. However, the use of short packets will inevitably bring performance loss, which is undoubtedly a huge challenge for IoT applications that require reliable and secure information transmission. Even worse, things get tricky as the classic Shannon's information theory is no longer applicable for short-packet communications\cite{durisi2016toward}, and using the existing information theory to analyze system performance may encounter inaccurate conclusions. Therefore, it has become an urgent problem to reconsider the analysis and design of reliable and secure transmissions for short-packet CSs.

\begin{table}[]
\centering
\caption{Notations}
\vspace{-3mm}
\resizebox{0.6\textwidth}{!}{
\begin{tabular}{ll}
\hline
$C$                      & : the capacity                                                                      \\
$C_s$                    & : the secrecy capacity                                                              \\
$C_b$                    & : the capacity of main channel (from Alice to Bob)                                  \\
$C_e$                    & : the capacity of wiretap channel (from Alice to Eve)                               \\
$R_b$                    & : the codeword rate                                                                 \\
$R_e$                    & : the rate redundancy                                                               \\
$R_s$                    & : the secrecy rate                                                                  \\
$B$                      & : the number of information bits per block                                          \\
$N$                      & : the blocklength (the number of channel uses)                                      \\
$\epsilon$               & : the decoding error probability                                                    \\
$\delta$                 & : the information leakage                                                           \\
$\bar\epsilon_A$           & : the avarage of $\epsilon$                                               \\
$\bar\epsilon$           & : the preset constraint of $\epsilon$                                               \\
$\bar\delta$             & : the preset constraint of $\delta$                                                 \\
$R^*(N,\epsilon)$        & : the maximum achievable channel coding rate                                        \\
$R^*(N,\epsilon,\delta)$ & : the maximal secret communication rate                                             \\
$\bar R_s$               & : the achievable secrecy rate (the achievability bound of $R^*(N,\epsilon,\delta)$) \\
$R_0$                    & : the coding rate                                                                   \\
$V$                      & : channel dispersion (in eq. (\ref{eq1}))                            \\
$V_1, V_2, V_3$          & : channel dispersion (in eq. (\ref{eq2}))                            \\
$\gamma_i,~i\in\{b,e\}$  & : the instantaneous SNR of the legitimate/eavesdropping channel                     \\
$\rho_i,~i\in\{b,e\}$    & : the average SNR of the legitimate/eavesdropping channel                           \\
$K, K_e$                 & : the number of antennas at Alice/Eve                                               \\
$M_1,M_2$                & : parameters used in \cite{yang2019}                               \\
$p_{out}$                & : the outage probability                                                            \\
$p_{so}$                 & : the secrecy outage probability                                                    \\
$\mu$                    & : the transmission threshold at Alice with the On-Off scheme                        \\
$\zeta$                  & : the preset limitation of $p_{out}$/$p_{so}$                                       \\
$\mathrm{E}\{\cdot\}$    & : the mean of the random variable $\{\cdot\}$                                       \\
$Q^{-1}\{\cdot\}$       & : the inverse of the Gaussian $Q$ function                                          \\
$\mathrm{P}\{\cdot\}$    & : probability of the event $\{\cdot\}$\\
\hline
\end{tabular}}
\vspace{-7mm}
\label{table1}
\end{table}
The article is organized as follows. We first outline the latest information theory progress with secure short-packet communications. Then ergodic/outage-based secrecy performance parameters and corresponding performance analysis frameworks for short-packet communications have been introduced. On this basis, recent works investigate diverse CSs with short packets from the perspective of PLS are reviewed. Subsequently, we discuss the challenges and future trends for secure transmission in IoT. Finally, conclusion of this article will be given. Table \ref{table1} lists the notations used in this article.

\section{Information Theory Progress for Short-Packet Communications}
\subsection{Reliable Short-packet Communications}
In classic information theory, capacity is usually used to measure the data-carrying capability, which is the maximum coding rate that an IFB system can support for error-free transmission. However, in the finite blocklength (FB) regime, the channel coding gain is severely reduced, and the error-free transmission cannot be guaranteed. Short-packet CSs are encountered with transmission reliability challenges.

In the past decade, there has been great interest in investigating the loss of data carrying capacity incurred by coding with FB, and the research on reliable short-packet CSs has made significant progress. According to \cite{Polyanskiy2010Channel}, the maximum achievable channel coding rate for a given blocklength $N$ with a constraint on the decoding error probability of $\epsilon$ at the receiver can be approximated as
\begin{align}
R^*(N,\epsilon)=C-\sqrt{\frac{V}{N}}Q^{-1}\left(\epsilon\right),
\label{eq1}
\end{align}
where $C$ is the capacity of IFB-CSs, $V$ is the channel dispersion that is determined by the channel state information (CSI) of the main channel from Alice to Bob. A penalty term proportional to $\frac{1}{\sqrt{N}}$ has been introduced in (\ref{eq1}) to characterize the performance loss caused by the use of FB. If $N$ approaches infinity, the penalty vanishes so that $R^*(N,\epsilon)$ converge to $C$. The above conclusion has been used to analyze the reliability performance of FB-CSs, such as two-way relaying and non-orthogonal multiple access (NOMA) system. However, neither of the above works considers secrecy performance.

\subsection{Secure Short-packet Communications}
\begin{figure}[!t]
\begin{center}
\includegraphics[width=3.5 in]{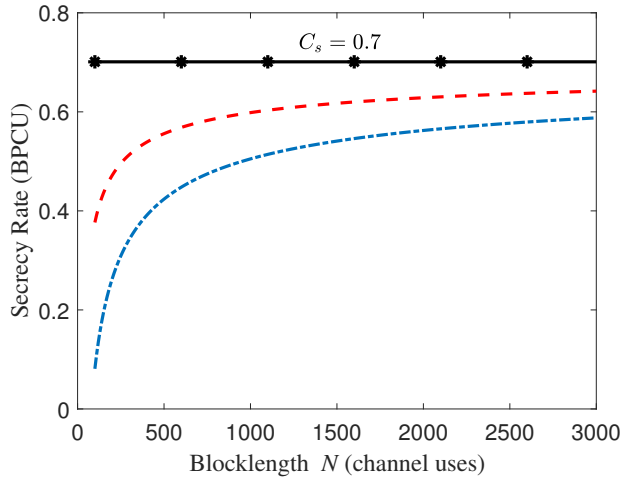}
\end{center}
\caption{\small{Secrecy rate for the Gaussian wiretap channel with $\gamma_b$ = 10dB, $\gamma_e$ =5dB, $\epsilon=\delta=10^{-3}$.}}
\label{fig2}
\vspace{-7mm}
\end{figure}

In IFB-CSs, the secrecy capacity $C_s$, formulated as the difference between the main channel capacity $C_b$ and the wiretap channel capacity $C_e$, is the basic feature to measure the secrecy performance. However, the above conclusion cannot be applied to evaluate the secrecy performance of short-packet CSs. By far, only limited works have studied the secrecy performance with FB from the information-theoretic perspective. Previous works are devoted to exploring achievability and converse boundaries for wiretap channels with security metrics and further improving the boundaries. The latest research shows that the achievability and converse bounds on the maximal secret communication rate $R^*(N,\epsilon,\delta)$ with a fixed blocklength $N$, decoding error probability $\epsilon$, and information leakage $\delta$ for general wiretap channels in \cite{Yang2017Wiretap}, which can be represented as
\begin{align}
C_s-\sqrt{\frac{V_1}{N}}Q^{-1}\left(\epsilon\right)-\sqrt{\frac{V_2}{N}}Q^{-1}\left(\delta\right)\stackrel{<}{\approx} R^*(N,\epsilon,\delta)\stackrel{<}{\approx}C_s-\sqrt{\frac{V_3}{N}}Q^{-1}\left(\epsilon+\delta\right),
\label{eq2}
\end{align}
where $V_1$, $V_2$, and $V_3$ are constants that depend on the instantaneous signal-noise ratio (SNR) of the legitimate and eavesdropping channel, denoted by $\gamma_b$ and $\gamma_e$ respectively. It can be observed in (\ref{eq2})  that \emph{perfect reliability and secrecy}\footnote{It means that the error probability at Bob and the information leakage for Eve can be arbitrarily small in the IFB domain.} cannot be guaranteed. Fig. \ref{fig2} plots $R^*(N,\epsilon,\delta)$ versus $N$ with given $\epsilon$ and $\delta$ in the FB domain. As the curves show, compared with the IFB case, the use of FB leads to a loss related to $N$, which diminishes as $N$ increases, and the upper and lower bounds of $R^*(N,\epsilon,\delta)$ coincide with $C_s$ asymptotically when $N\rightarrow\infty$.

However, applying the above-mentioned conclusion of secure short-packet information theory to actual systems for performance analysis still encounters challenges:
\begin{itemize}
\item
Compared with $C_s$ applicable to IFB systems, the use of FB incurs a penalty term related to blocklength $N$, making the form of $R^*(N,\epsilon,\delta)$ too complicated to be applied for performance analysis.
\end{itemize}
\begin{itemize}
\item
Besides, an accurate and unique $C_s$ can be obtained for a given channel realization when IFB coding scheme is adopted. However, only boundaries of $R^*(N,\epsilon,\delta)$ can be obtained in the FB domain, which inevitably brings deviations in performance analysis.
\end{itemize}
\begin{itemize}
\item
What's more, the difference in the information theory result calls for new coding scheme. For the IBF regime,  the Wyner's wiretap coding scheme can be used to design the codeword rate $R_b$ and rate redundancy $R_e$, respectively, so that the secrecy rate $R_s$ is as close as possible to $C_s$. However, the Wyner's scheme is no longer applicable for FB-CSs, we should design coding rate $R_0$ directly to make it close to the achievable secrecy rate $\bar R_s$\footnote{To obtain a conservative performance evaluation, we use the lower bound on $R^*(N,\epsilon,\delta)$, the achievable secrecy rate $\bar R_s$, for performance analysis.}.
\end{itemize}

Below, based on above observations, we provide performance evaluation parameters and corresponding analysis frameworks for short-packet CSs with the conclusion of secure short-packet information theory.
\section{Secrecy Performance Metrics in Secure Short-Packet Communications}
\label{III}
Information theory mainly examines the secrecy capacity (maximal secret communication rate in the FB domain) performance under given channel states. However, in actual communication scenarios, it is often accompanied by factors such as fading, multi-path, and shadow that may cause channel time-varying. It is necessary to establish secrecy performance metrics that can be applied to performance analysis of actual systems. Typical performance metrics can be broadly divided into two categories, ergodic-based and outage-based. Specifically, ergodic-based performance metrics refers to the time average of performance in all fading states, while outage-based metrics allow to evaluate system performance under a certain restriction of outage probability. In this part, we will introduce the ergodic-based and outage-based performance metric that are suitable for evaluating short-packet CSs, respectively.

\subsection{Ergodic-based Performance Metric}
\label{Ergodic-based}
To analyze the performance of actual short-packet CSs, the authors in \cite{yang2019} define the secrecy throughput, i.e., the average transmission rate at which data packets can be transmitted reliably under a certain preset secrecy constraint $\bar\delta$, as the performance metric. Specifically, assume that the Access Point (AP) transmits $B$ information bits over $N$ channel uses for each sporadic short-packet transmission, that is, a fixed $R_0=\frac{B}{N}$ is adopted for the entire transmission process. By reformulating the achievable secrecy rate $\bar R_s$, the decoding error probability with transmission rate $R_0$ under given CSI and preset secrecy constraint $\bar\delta$ can be characterized by $\epsilon(\gamma_b,\gamma_e,\bar\delta, R_0)$, then the secrecy throughput can be formulated as $T=R_0(1-\bar\epsilon_A)$, where $\bar\epsilon_A=\mathrm{E}_{\gamma_b,\gamma_e}\left[\epsilon(\gamma_b,\gamma_e,\bar\delta, R_0)\right]$ is the average decoding error probability.

It is worth noting that an analytical framework is proposed in \cite{yang2019} to approximate the average secrecy throughput of short-packet CSs, and closed-form approximations for the secrecy throughput have been derived for both cases in which the AP has one antenna or multiple antennas with the proposed analytical framework. By solving the optimization problem of maximizing the average secrecy throughput, $R_0$ for the entire transmission process can be obtained. Furthermore, the influence of system parameters on the trade-off between transmission delay and reliability under secrecy constraints is observed in \cite{yang2019}.

\begin{figure}[!t]
\begin{center}
\includegraphics[width=3.5 in]{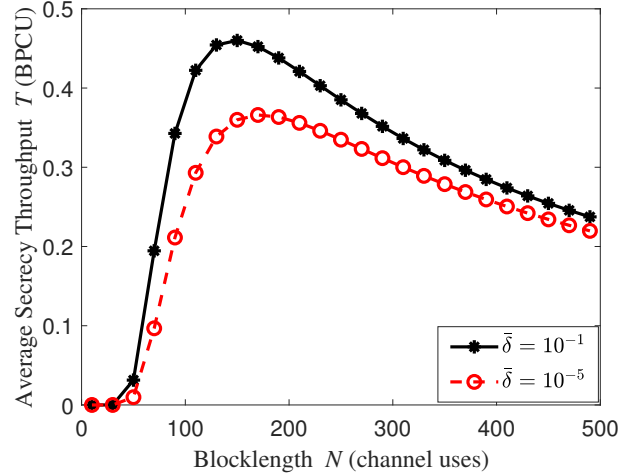}
\end{center}
\caption{\small{ The average secrecy throughput $T$ under the single-antenna case versus blocklength $N$ with system parameters:  $K_e$ = 2, $\rho_b$ = 10 dB, $\rho_e$ = 3 dB, $B$ = 200 bits, $M_1$ = 10, and $M_2$ = 20.}}
\label{fig3}
\vspace{-7mm}
\end{figure}

Fig. \ref{fig3} depicts the average secrecy throughput $T$ obtained by the analytical result versus $N$ and $\bar\delta$ under the case with a single-antenna AP. It can be observed in Fig. \ref{fig3} that $T$ increases first and then decreases as $N$ increases. The reason behind this is that $R_0$ and  $\bar\epsilon_A$ decrease as $N$ increases, which reflects the trade-off between transmission performance and transmission rate. There exists an optimal $N$ to maximize $T$, which can be obtained by the analysis framework provided by \cite{yang2019}. Moreover, it can be seen that $T$ decreases with $\bar\delta$, which indicates that a relaxed secrecy limitation will result in a greater throughput.

The performance parameters and analysis framework proposed in \cite{yang2019} have been extended to the short-packet communication scenario with multiple eavesdroppers by \cite{2020throughputMultiEve}. In this work, the impact of the number of eavesdroppers on the security performance of a short-packet system is investigated. The evaluation in \cite{2020throughputMultiEve} shows that the average secrecy output decreases with the number of eavesdroppers, resulting in a lower average secrecy throughput. The above-mentioned works have established corresponding performance evaluation parameters and analysis frameworks with the conclusion of secure short-packet information theory, and provide feasible solutions for secure performance analysis of actual FB-CSs. However, considering that there exist a decoding error probability and an information leakage for short-packet communication, as well as the inability to ensure that the transmission can always meet the reliability and security constraints, it is more appropriate to adopt outage-based secrecy performance metrics.
\subsection{Outage-based Performance Metric}
In information theory, secure communication relies on the advantage of main channel over eavesdropping channel. Nevertheless, due to the fluctuation of the fading channel, there is always a probability that the quality of the main channel is worse than that of the eavesdropping channel. If it occurs, the secure transmission of confidential information cannot be guaranteed. To measure the probability of this occasion happening, we introduce the concept of secrecy outage probability (SOP). On this basis, outage-based performance metrics have been introduced to evaluate the secrecy performance of CSs.

For previous IFB-CSs, the basic SOP definition can be represented as $p_{out}=\mathrm{Pr}\{C_s<R_s\}$, which gives the feature of satisfying both reliable and secure transmission. However, it cannot be used for FB-CSs as there exist a decoding error probability and an information leakage. In \cite{feng2021SOP}, we establish a definition of outage probability considering \emph{reliability and security simultaneously} to evaluate system performance based on the characteristics of short-packet transmission. Specifically, the outage event has been defined as the case that the transmission with current $R_0$ and $N$ violates either a preset reliable constraint $\bar\epsilon$ or a preset secure constraint $\bar\delta$, with a probability $p_{out}=\mathrm{Pr}\{\bar R_s(N,\bar\epsilon,\bar\delta)\leq R_0\}$. On this basis, the effective throughput $T$ is establish to measure the average effectively received information bits per channel use subject to an outage constraint $p_{out}<\zeta$, where $\zeta$ is a pre-established threshold of outage probability. Notably, the \emph{effective transmission} refers to the transmission that satisfies both the reliable and secure constraints.

To cope with the complicated form of secure short-packet information theory result, a general analysis framework based on distribution approximation is proposed to obtain the approximations of the outage probability and effective throughput. Specially, closed-form expressions for these quantities are derived for the high SNR regime. The optimal coding rate for the whole transmission that maximizes the effective throughput can be obtained by searching method.

\begin{figure}[!t]
\begin{center}
\includegraphics[width=3.5 in]{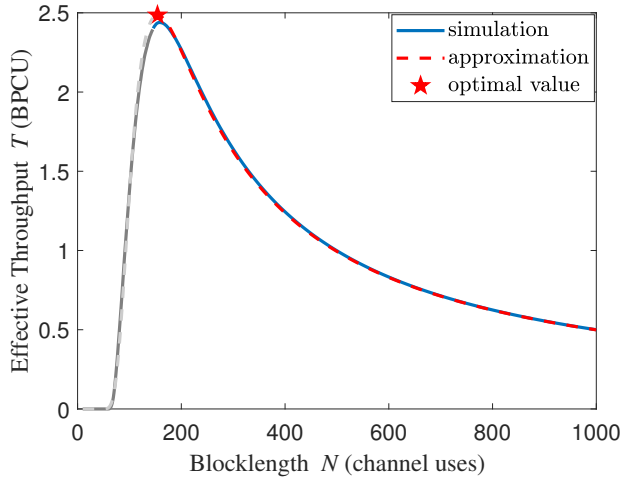}
\end{center}
\caption{\small{ The effective throughput $T$  versus blocklength $N$ with system parameters:  $K$ = 8, $\rho_b$ = 15 dB, $\rho_e$ = 10 dB, $B$ = 500 bits, $\bar\epsilon=\bar\delta=10^{-3}$ and $\zeta=0.3$.}}
\label{fig4}
\vspace{-7mm}
\end{figure}

Fig. \ref{fig4} plots the effective throughput $T$ obtained by simulation and approximation versus $N$. It can be seen that the approximate results fit well with the simulation results, and the optimal value obtained by analysis is close to the true value, which have verified the feasibility and accuracy of the proposed framework. It should be noted that the value range of $N$ corresponding to the gray curve in Fig. \ref{fig4} cannot meet the outage restriction $p_{out}<\zeta$. Similarly, it can be observed in Fig. \ref{fig4} that the choice of $N$ actually implies a tradeoff between transmission rate and reliable-secure performance. With the increase of $N$, the probability of effective transmission $(1-p_{out})$ increases while transmission rate $R_0$ decreases, further $T$ increases first and then decreases due to combined effects of these two.

Nevertheless, the definitions mentioned above fails to distinguish between reliability outage (violate reliable restriction) and secrecy outage (violate secure restriction). This problem is solved in \cite{Zhou2011Rethinking} by redefining the SOP on the IFB domain. Specifically, In \cite{Zhou2011Rethinking}, an on-off transmission scheme has been employed at transmitter to ensure that confidential information can be transmitted under the premise of satisfying the reliable restriction $R_b<C_b$. Then an outage only occurs when $C_e> R_e$. The corresponding SOP can be redefined as $P_{so}=P(C_e>R_e|\text{message transmission})$, conditioned upon a message actually being transmitted. This definition provides a more explicit measure of the level of security if the instantaneous main channel CSI is available at Alice.

Inspired by this work, an evolved SOP definition for short-packet CSs can be proposed. Specifically, a similar on-off transmission scheme based on the main channel CSI can be adopted at transmitter, and the corresponding transmission threshold $\gamma_b>\mu$ is set so that the system can only transmit information when the reliability condition $R_0<R^*(N,\epsilon,\gamma_b)$ is met.
Since Alice cannot obtain the instantaneous wiretap channel CSI, which varies randomly with time, the current achievable secrecy rate $\bar R_s(\gamma_b,\gamma_e)$ cannot be calculated. Therefore, transmitting information at a coding rate $R_0$ under the condition $\gamma_b>\mu$ will encounter two situations:

1)$R_0\leq\bar R_s$, transmission with a coding rate $R_0$ can satisfy the preset reliability constraint $\bar\epsilon$ and the preset secrecy constraint $\bar\delta$, simultaneously.

2)$R_0>\bar R_s$, the reliability condition $R_0<R^*(N,\epsilon,\gamma_b)$ is still satisfied while the information leakage will exceed the preset $\bar\delta$, that is, secrecy cannot be guaranteed. This situation is defined as a \emph{secrecy outage event}, with a probability $p_{so}(\gamma_b,\bar\epsilon,\bar\delta)=\mathrm{Pr}\{R_0>\bar R_s|\gamma_b>\mu\}$.

There are two significant differences between the SOP definition proposed for short-packet systems with the one in  \cite{Zhou2011Rethinking}. For FB-CSs, $\bar R_s$ is used to replace $C_s$, resulting in an upper bound of the true SOP essentially. On the other hand, the secrecy outage event in the previous definition refers to the breach of the perfect secrecy, which is interpreted as the information leakage exceeds $\bar\delta$ with FB.

\begin{figure}[!t]
\begin{center}
\includegraphics[width=3.5 in]{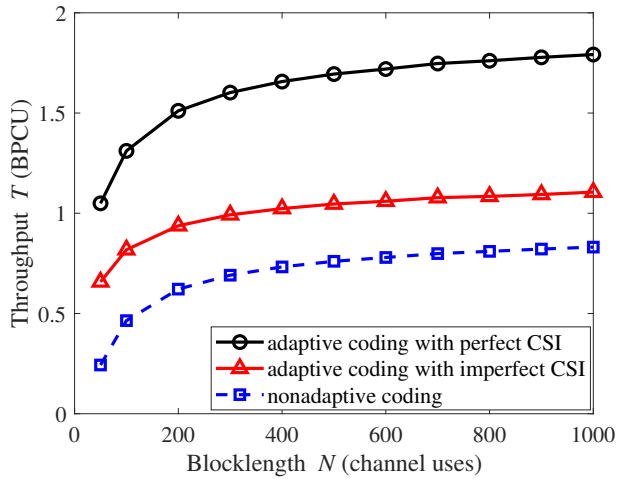}
\end{center}
\caption{\small{ The throughput $T$ versus blocklength $N$ with system parameters:  $K$ = 8, $\rho_b$ = 15 dB, $\rho_e$ = 10 dB, $B$ = 500 bits, $\bar\epsilon=\bar\delta=10^{-3}$ and $\zeta=0.3$.}}
\label{fig5}
\vspace{-7mm}
\end{figure}
Further, the reliable throughput $T$ can be defined as the transmission rate satisfying SOP constraint $p_{so}<\zeta$. On this basis, adaptive/non-adaptive coding schemes can be designed separately according to the explicit/partial instantaneous CSI of the main channel. Corresponding design strategy of scheme parameters such as transmission threshold and coding rate can be obtained by solving the optimization problem of maximizing throughput.

Specifically,  Fig. \ref{fig5} depicts the throughput obtained by adaptive/nonadaptive scheme versus $N$. Obviously, the throughput under the adaptive coding scheme is significantly greater than that of the non-adaptive coding scheme, as the former makes full use of CSI. Fig. \ref{fig5} also shows the impact of the imperfect CSI caused by the channel information quantization error on the secrecy performance with the adaptive encoder scheme. Obviously, from Fig. \ref{fig5}, it can be acquired that channel estimation imperfection will lead to the deterioration of system security performance. Notably, the non-adaptive coding scheme can essentially be regarded as the worst case with imperfect CSI, which quantizes and encodes the main channel CSI to obtain only $1$bit information for channel feedback. Nevertheless, the non-adaptive coding scheme still has application value due to its low complexity and support for offline design of transmission parameters.

\section{Application of physical layer security in short-packet communication systems}
With the aid of the the latest results in \cite{Yang2017Wiretap}, as well as performance parameters and corresponding analysis frameworks mentioned earlier, from the perspective of PLS, many recent researches have analyzed short-packet CSs under different communication scenarios or adopting different security strategies.

To meet transmission requirements and objective challenges of IoT applications, some specific communication scenarios have been proposed. Recently, the secrecy performance of these communication systems on the FB domain have been studied. The authors in \cite{2020CognitiveIoT} focus on the secure short-packet communication in cognitive IoT\footnote{Cognitive radio is a technology that efficiently utilizes spectrum resources, the IoT network that adopts this technology is called cognitive IoT. }. A cognitive IoT network in the presence of a passive eavesdropper containing a pair of main transceivers and a pair of auxiliary transceivers that share spectrum resources is considered.  The corresponding secrecy throughput is obtained by adopting the analytical framework proposed in \cite{yang2019}. In \cite{2020UAV}, secure short-packet communication is extended to the CSs with unmanned aerial vehicles (UAV)\footnote{UAV can be used as an air base station or relay node to cope with the lack of infrastructure in several IoT systems.}. An approximate version of system average effective security rate can be obtained, which can be maximized by jointly designing the UAV's trajectory and transmit power with the constraints of UAV's mobility and transmit power. To address the above-mentioned non-convex optimization problem, the authors in \cite{2020UAV} propose a corresponding analysis framework and develop an alternating iterative algorithm based on continuous convex approximation (SCA) technology.

In addition, some researchers are concerned about the impact of PLS strategies on improving the secrecy performance of short-packet CSs. Multi-antenna technology is considered in \cite{2020UplinkMU-MIMO}. Specifically, a multi-user multiple-input multiple-output (MU-MIMO) system is studied, which is composed of a base station and an eavesdropping node equipped with multiple antennas, respectively, as well as multiple single-antenna user devices. The system average secrecy throughput is derived to measure the secrecy performance. The simulation results show that appropriately increasing the number of antennas at the BS and the transmission SNR can effectively improve system secrecy performance. The choice of transmission blocklength will also affect the security performance.
In \cite{2020NOMA-Assisted}, it is claimed that NOMA can be utilized as a lightweight strategy for improving PLS performance without introducing extra security mechanisms. Both downlink and uplink NOMA schemes are designed for secure transmission in \cite{2020NOMA-Assisted}. What's more, numerical results confirm that the use of NOMA scheme will improve the secrecy performance compared with OMA scheme. Further, the authors in \cite{2020NOMABLER} analyze the secure performance of the NOMA downlink short-packet CSs in a flat Rayleigh fading channel from the perspective of average secure block error rate (BLER). In \cite{2020NOMABLER}, an entrusted central user and an untrusted cell-edge user are considered, and NOMA protocol is employed by the base station to send confidential messages to the central user and cell-edge user, respectively.  By utilizing the linear approximations on BLER and secure BLER, derived from the information theory results shown in \cite{Polyanskiy2010Channel} and \cite{Yang2017Wiretap}, the average secure BLER of the central user has been deduced, which has been used to observe some insights about the impact of main system parameters on the secure performance of the NOMA-assisted systems.

Some researchers are also devoted to exploring the resource allocation problem of secure short-packet CS in a multi-user scenario. The reference \cite{Ren2020ResourceAllocationTCOM} is the first work in this field. In \cite{Ren2020ResourceAllocationTCOM}, the weighted sum throughput (WST) maximization problem under the restriction of total power and total channel blocklength, as well as the total transmit power (TTP) minimization problem while guaranteeing the minimum security capacity of each device and the total channel blocklength, have been solved by optimizing the power and channel bandwidth unit allocation jointly. The corresponding algorithms are proposed to solve the above two optimization problems, respectively. The TTP problem under the multi-antenna AP system with artificial noise scheme has been further explored in \cite{Ren2020ResourceAllocationICC}. A low-complexity algorithm has been propose in \cite{Ren2020ResourceAllocationICC} to find a sub-optimal solution for this problem, which ensures the secrecy of the uRLLC users transmissions while yields significant power savings.

\section{Future research directions and challenges}
\subsection{The Influence of Channel Estimation on the Delay-Transmission Efficiency Trade-off}
In short-packet CSs, the choice of transmission scheme depends largely on the prior knowledge of CSI, which will affect the system performance directly. However, the accuracy of channel estimation will be restricted by other factors in an IoT system. Specifically, due to the use of short packets, accurate channel estimation will increase additional overhead, thereby reducing the transmission efficiency of data information significantly. On the contrary, if a rough channel estimation is applied, the CSI will not be fully utilized to design a flexible coding scheme, resulting in a reduction in transmission rate and transmission efficiency.
In addition, delay will be introduced by the channel estimation and feedback process, which is not suitable for delay-sensitive application scenarios. How to design an accurate and effective channel estimation method for short-packet CSs to achieve the best balance between delay requirements and transmission efficiency has become an urgent problem to be solved. Furthermore, how to maximize the use of known CSI and reduce the impact of imperfect CSI on transmission performance remains to be further explored.

\subsection{Resource Allocation for Multi-user System}
IoT systems often involve massive resource-constrained IoT nodes, so reasonable resource allocation is essential.Recent work has initially explored the resource allocation problem of secure short-packet CSs in multi-user scenarios\cite{Ren2020ResourceAllocationICC,Ren2020ResourceAllocationTCOM}. In the future, the resource allocation problem can be extended to short-packet CSs deployed with PLS scheme, corresponding optimization problems should be established based on the characteristics and requirements of the transmission scheme. Furthermore, low-complexity algorithms should be explored to solve optimization problems, so as to provide theoretical guidance for resource allocation in short-packet CSs. In addition, the application of machine learning methods in resource allocation problems is worthy of further exploration.

\subsection{Lightweight PLS Scheme}
Considering that the IoT system is resource-constrained, lightweight secrecy strategies need to be adopted. In this article, NOMA has been used as an example to discuss the impact of low-complexity security schemes on short-packet CSs. Subsequent research can further evaluate the performance of secure strategies such as on-off transmission, space-time coding and cooperative communication based on the performance parameters and analysis framework mentioned in this article, then select promising secure solutions from them for potential 6G applications. What's more, combining emerging 6G technologies such as intelligent reflecting surface to improve system secrecy performance should also be given full attention. In addition, how to design secrecy coding schemes for short-packet CSs to reach the achievable secrecy rate as close as possible is still an open issue.

\section{Conclusions}
In this article, we provide an overview of the latest research results of secure short-packet wireless communication from the perspective of PLS. On the basis of reviewing the state-of-the-art progress of secure short-packet information theory, this work provides ergodic/outage-based secrecy performance metrics which are suitable for performance analysis of short-packet systems. Furthermore, recent works investigate diverse CSs with short packets from the perspective of PLS are reviewed. Some valuable research directions on the topic of secure short-packet communication have been provided. We hope this overview can provide help and inspiration for subsequent research and promote more important and practical research in 6G.

\renewcommand{\baselinestretch}{1.1}

\end{document}